\documentclass{llncs}
\usepackage{amsmath}
\usepackage{amsfonts}
\usepackage{floatflt}
\usepackage{float}
\usepackage{graphicx}

\input{macros.sty}

\title{Comparison of Algorithms for Checking Emptiness on B{\"u}chi Automata}
\author{Andreas Gaiser\inst{1}\thanks{The author was supported by the DFG Graduiertenkolleg 1480 (PUMA).} \and Stefan Schwoon\inst{2}}
\institute{
Institut f{\"u}r Informatik,
Technische Universit{\"a}t M{\"u}nchen, Germany
\and
LSV, CNRS, ENS de Cachan, INRIA Saclay, France \\
\email{gaiser@model.in.tum.de, schwoon@lsv.ens-cachan.fr}}

\begin{document}
\sloppy
\maketitle

\begin{abstract}
We re-investigate the problem of LTL model-checking for finite-state
systems. Typical solutions, like in Spin,
work on the fly, reducing the problem to B\"uchi emptiness.
This can be done in linear
time, and a variety of algorithms with this property exist. Nonetheless,
subtle design decisions can make a great difference to their actual
performance in practice, especially when used on-the-fly.
We compare a number of algorithms experimentally on a large benchmark suite,
measure their actual run-time performance, and propose improvements.
Compared with
the algorithm implemented in Spin, our best algorithm is faster
by about 33~\% on average.
We therefore recommend that, for on-the-fly explicit-state
model checking, nested DFS should be replaced by better solutions.
An abridged version of this paper has appeared in~\cite{GS09}.
\end{abstract}

\section{Introduction}
\label{sec:intro}

Model checking is the problem of determining whether a given hardware or
software system meets its specification.
In the automata-theoretic approach,
the system may have finitely many states, and
the specification is an LTL formula, which is
translated into a B\"uchi automaton,
intersected with the system,
and checked for emptiness. Thus, model checking becomes
a graph-theoretic problem.

Because of its importance, the problem has been intensively investigated.
For instance, \emph{symbolic} algorithms use efficient data structures
such as BDDs to work on sets of states; a
survey of them can be found in~\cite{FFK+01}. Moreover, 
\emph{parallel} model-checking algorithms have been developed~\cite{BBR08}.
The best known symbolic or parallel solutions have suboptimal
asymptotic complexity ($\mathcal{O}(n\log n)$, where
$n$ is the number of states), but are often faster than that in practice.

B\"uchi emptiness can also be solved in $\mathcal{O}(n)$ time.
All known linear algorithms are
\emph{explicit}, i.e.\ they construct and explore states one by one,
by depth-first search (DFS).
Typically, they compute some data about each state:
its unique \emph{state descriptor} 
and some \emph{auxiliary data} needed for the
emptiness check. Since the state descriptor is usually much larger than the
auxiliary data, approximative techniques such as bitstate hashing have been
developed that avoid them,
storing just the auxiliary information in a hash table~\cite{HPY96}. This
entails the risk of undetectable hash collisions; however
the probability of a wrong result can 
be reduced below a chosen threshold by repeating the emptiness test
with different hash functions. Thus they represent
a trade-off between time and memory requirements.
Henceforth, we shall refer to
non-approximative methods that do use state descriptors as \emph{exact}
methods.

We further identify two subgroups of explicit algorithms:
\emph{Nested-DFS} methods
directly look for acceptings cycle in a B\"uchi automaton;
they need very little auxiliary memory and work well with bitstate hashing.
\emph{SCC-based}
algorithms identify strongly connected components containing accepting cycles;
they require more auxiliary memory
but can find counterexamples more quickly.

All explicit algorithms can work ``on-the-fly'', i.e.\ the
(intersected) B\"uchi automaton is not known at the outset.
Rather, one begins with a B\"uchi automaton for the formula (typically small)
and a compact system description and extracts
the initial state from these.
Successor states are computed during exploration as needed.
If non-emptiness is detected, the algorithms terminate before constructing
the entire intersection. Moreover, in this approach the
transition relation need not be stored in memory. As we shall see, the
on-the-fly nature of explicit algorithms is very significant when evaluating
their performance properly.

In this paper, we investigate performance aspects of explicit,
exact, on-the-fly algorithms for B\"uchi emptiness. The best-known
example for such a tool is Spin~\cite{Hol03}, which
uses the nested-DFS algorithm proposed by Holzmann et al~\cite{HPY96},
henceforth called HPY.
The reasons for this choice are partly historic; the faster detection
capabilities of SCC-based algorithm were not known when Spin was designed,
having first been pointed out by Couvreur in 1999~\cite{Cou99}. Thus, the
status of
HPY as the best choice is questionable, all the more so
since the memory advantages of nested DFS are comparatively scant in our
setting. Moreover,
improved nested DFS algorithms have been proposed in the meantime.

We therefore evaluate several algorithms
based on their actual running time and memory usage
on a large suite of benchmarks. Previous papers, especially those
on SCC-based algorithms~\cite{GV04,SE05,CDP05,GV05}, provided similar
experimental results, however, experiments were few or random 
and unsatisfying
in one important aspect: they worked from pre-computed B\"uchi automata,
rather than truly on-the-fly. This aspect will play
a significant role in our evaluation.

To summarize, this paper contains the following contributions and findings:
\begin{itemize}
\item We provide improvements in both subgroups,
  nested DFS and SCC-based. These concern the algorithms
  of Couvreur~\cite{Cou99} and Schwoon/Esparza~\cite{SE05}.
\item One of the algorithms we study can be extended to generalized B\"uchi
  automata, and we investigate this aspect.
\item We implemented existing and new algorithms and compare them
  on a large benchmark suite. We analyze the structural properties of
  B\"uchi automata that cause performance differences.
\end{itemize}

We make the following observations:
The overall memory consumption of all algorithms is dominated by
the state descriptors, the differences in auxiliary memory play
virtually no role. The running times
depend practically exclusively on the number of successor
computations. When experimenting with pre-computed
automata -- as done in some other papers -- this operation becomes cheap,
which causes misleading results.
Our results allow to derive
detailed recommendations which algorithms to use in which circumstances.
These recommendations revise those from~\cite{SE05}; Couvreur's algorithm
which was recommended there, is shown to have weak performance; however,
the modification mentioned above amends it. Moreover, our modification of
Schwoon/Esparza improves the previous best nested-DFS algorithm.

In addition, this paper provides new, self-contained proofs of both
improved algorithms. Since the original algorithms are already known
to be correct,
one could easily give non-self-contained proofs by showing that
the modifications do not affect correctness. However, we feel that
there are still good reasons to provide completely new proofs.

First, the nested-DFS algorithm was derived through a succession of
modifications, from~\cite{CVWY92} via~\cite{HPY96},
\cite{GMZ04}, and~\cite{SE05}, during which the mechanics of
the algorithm have changed sufficiently to merit a new proof.

Secondly, self-contained proofs are a necessity if
improved B\"uchi emptiness algorithms are ever to be taught in
verification classes. In the authors' experience, DFS algorithms
are notoriously difficult to explain, yet the proofs we give are
still reasonably simple. For instance, the proof of the new SCC-based
algorithm is based on
eight simple facts that are easy to understand and prove. In our
experience, these proofs can be used in a classroom setting even
if the students are previously unfamiliar with the concepts of
DFS and SCCs.

We proceed as follows: Section~\ref{sec:buechi} establishes preliminaries,
Sections~\ref{sec:nested} and~\ref{sec:scc} present
nested-DFS and SCC-based algorithms, including our modifications.
Section~\ref{sec:experiments} details our experimental results and concludes.

\section{Preliminaries}
\label{sec:buechi}

A \emph{B\"uchi automaton} (BA) is a tuple $\B=(S,s_I,\post,A)$, where
$S$ is a finite set of \emph{states},
$s_I\in S$ is the \emph{initial state},
$\post\colon S\to2^S$ is the \emph{successor function}, and
$A\subseteq S$ are the \emph{accepting states}.
A \emph{path} of $\B$ is a sequence of states $s_1\cdots s_m$ for some $m\ge1$
such that $s_{i+1}\in\post(s_i)$ for all $1\le i<m$. If a path from $s$ to $t$
exists, we write $s\reach t$.
When $m>1$, we write $s\reachplus t$, and if additionally
$s=t$, we call the path a \emph{loop}.
A \emph{run} of $\B$ is an infinite sequence $(s_i)_{i\ge0}$
such that $s_0=s_I$ and $s_{i+1}\in\post(s_i)$ for all $i\ge0$. A run
is called \emph{accepting} if $s_i\in A$ for infinitely many different~$i$.
The \emph{emptiness problem} is to determine whether no accepting run
exists. If an accepting run exists, it is also called a \emph{counterexample}.
From now on, we assume a fixed B\"uchi automaton~$\B$.

Note that we omit the usual input alphabet because we are just interested in
emptiness checks. Moreover,
the transition relation is given as a mapping from each state to its
successors, which is suitable for on-the-fly algorithms.

A \emph{strongly connected component} (SCC) of $\B$ is a subset $C\subset S$
such that for each pair $s,t\in C$, we have $s\reach t$, and moreover,
no other state can be added to $C$ without violating this property.
An SCC~$C$ is called \emph{trivial} if $|C|=1$ and for the singleton
$s\in C$, \ $s\notin\post(s)$.
The following two facts are well-known:
\begin{itemize}
\item[(1)] A counterexample exists iff there exists some $s\in A$ such that
  $s_I\reach s$ and $s\reachplus s$. This fact is exploited by nested-DFS
  algorithms.
\item[(2)] A counterexample exists iff there exists a non-trivial SCC~$C$
  reachable from $s_I$ such that $C\cap A\ne\emptyset$.
  This fact is exploited by SCC-based algorithms.
\end{itemize}

A B\"uchi automaton is called \emph{weak} if each of its SCCs is either
contained in $A$ or in $S\setminus A$. This implies the following fact:
\begin{itemize}
\item[(3)]
Each loop in a weak BA is entirely contained in $A$ or in $S\setminus A$.
\end{itemize}

A \emph{generalized B\"uchi automaton} (GBA) is a tuple
$\G=(S,s_I,\post,\A)$, where $S$, $s_I$, and $\post$ are as before,
and $\A=(A_1,\ldots,A_k)$ is a \emph{set} of acceptance conditions,
i.e.\ $A_j\subseteq S$ for all $j=1,\ldots,k$. Paths and runs are
defined as for normal B\"uchi automata; a run $(s_i)_{i\ge0}$ of $\G$ is
called \emph{accepting} iff for each $j=1,\ldots,k$ there exist
infinitely many different $i$ such that $s_i\in A_j$.

GBA are generally more concise than BA:
a GBA with $k$ acceptance conditions and $n$ states can be
transformed into a BA with $nk$ states. There is no known nested-DFS algorithm
that avoids this $k$-fold blowup for checking emptiness of a GBA, although
Tauriainen's algorithm mitigates it~\cite{Tau06}. Some SCC-based algorithms,
however, can exploit the following fact:
\begin{itemize}
\item[(4)] A counterexample exists in $\G$ iff there exists a non-trivial
  SCC~$C$ reachable from $s_I$
  such that $C\cap A_j\ne\emptyset$ for all $j=1,\ldots,k$.
\end{itemize}

\section{Nested depth-first search}
\label{sec:nested}

Nested DFS was first proposed by Courcoubetis et al~\cite{CVWY92},
and all other 
algorithms in this subgroup still follow the same pattern.
There are two DFS iterations: the ``blue'' DFS is the
main loop and marks
every newly discovered state as blue. Upon backtracing
from an accepting state~$s$, it initiates a ``red'' DFS that tries to find
a loop back to $s$, marking every encountered state as red.
If a loop is found, a counterexample is reported, otherwise the blue DFS
continues, but the established red markings remain. Thus, both blue and red
DFS visit each state at most once each. Only two bits of auxiliary data are
required per state.

This pattern of searching for accepting loops in post-order ensures
that multiple red searches do not interfere; states in ``deep'' SCCs
are coloured red first, and when a red DFS terminates, red states are guaranteed
not to be part of any counterexample. While being memory-efficient and simple,
this has two disadvantages. First, nested DFS prefers long
counterexamples over shorter ones; secondly, the blue DFS never notices that
a complete counterexample has already been explored and continues exploring
potentially many more states than necessary before eventually noticing the
counterexample during backtracking. Also, nested DFS computes the successors
of many states twice.

Several improvements have been suggested in the past, e.g.\ the 
HPY algorithm~\cite{HPY96}, implemented in Spin, and the SE
algorithm~\cite{SE05}. We present an improvement of SE, shown in
Figure~\ref{fig:newnested}.
We first describe the differences w.r.t.\ SE; a detailed proof
is given below.

The additions to SE are in lines~\ref{li:allredtrue}
and from \ref{li:allredfalse} to \ref{li:makered}. These exploit the fact
that red states cannot be part of any counterexample; therefore a state
that has only red successors cannot be either. This avoids certain
initiations of the red search. The improvement is similar in spirit
to~\cite{GMZ04}, but avoids some
unnecessary invocations of $\post$. Like in~\cite{CVWY92},
only two bits per state are used. Our experiments shall
show that it performs best among the known nested DFS algorithms.


\begin{figure}[t]
\begin{center}
\begin{minipage}[t]{6.5cm}
{\algo
[procedure] {\it new\_dfs} ()
>[call] {\it dfs\_blue}($s_I$)}			\label{li:bluecall1}

\bigskip

{\algcontd
[procedure] {\it dfs\_blue} ($s$)
>{\it allred} := [true];  			\label{li:allredtrue}
>$s.{\it colour} := {\it cyan}$;		\label{li:makecyan}
>\for{t\in\post(s)}
>>[if] $t.{\it colour} = {\it cyan}$
>>>>$\land\ (s\in A\lor t\in A)$ [then]
>>>[report cycle]				\label{li:bluecycle}
>>[else] \ifthen{t.{\it colour}={\it white}}
>>>[call] {\it dfs\_blue}(t);			\label{li:bluecall2}
>>[if] $t.{\it colour} \ne {\it red}$ [then]	\label{li:allredfalse}
>>>{\it allred} := [false];}
\end{minipage}
\begin{minipage}[t]{5.5cm}
{\algcontd
>[if] $\it allred$ [then]			\label{li:ifallred}
>>$s.{\it colour}$ := {\it red}			\label{li:makered}
>[else] [if] $s\in A$ [then] 			\label{li:ifaccept}
>>[call] {\it dfs\_red}(s);			\label{li:redcall}
>>$s.{\it colour} := {\it red}$			\label{li:makered3}
>[else]
>>$s.{\it colour} := {\it blue}$		\label{li:makeblue}}

\smallskip

{\algcontd
[procedure] {\it dfs\_red} ($s$)
>\for{t\in\post(s)}
>>\ifthen{t.{\it colour} = {\it cyan}}		\label{li:redfound}
>>>[report cycle]				\label{li:redcycle}
>>[else] \ifthen{t.{\it colour}={\it blue}}
>>>$t.{\it colour} := {\it red}$;		\label{li:makered2}
>>>[call] {\it dfs\_red}(t)}
\end{minipage}
\end{center}
\caption{New Nested-DFS algorithm.}
\label{fig:newnested}
\end{figure}

Finally, we remark that for weak automata a much simpler
algorithm suffices, as observed by \v{C}ern\'a and
Pel\'anek~\cite{CP03}. Exploiting Fact~(3),
one can simply omit the red search because all counterexamples are bound
to be reported by line~\ref{li:bluecycle} in Figure~\ref{fig:newnested}. In
that case, $\post$ is only invoked once per state.

\subsection{Proof of the new algorithm}

\def\dfsblue{{\it dfs\_blue}}
\def\dfsred{{\it dfs\_red}}

\paragraph{Colour changes}
We assume that all newly discovered states are initialized to white.
There are four colours, meaning that the auxiliary data can be encoded
with two bits. There are five statements that change the colour of
states, in lines \ref{li:makecyan}, \ref{li:makered}, \ref{li:makered3},
\ref{li:makeblue}, and~\ref{li:makered2}

The procedure \dfsblue{} is only invoked on white states in
lines \ref{li:bluecall1} and~\ref{li:bluecall2}. Thus, the statement
in line 5 changes only white states into cyan. There is no statement
that changes states back to white, therefore \dfsblue{} is only
invoked once per state. The statement in line~\ref{li:makered2}
changes only blue states to red. Therefore, when \dfsblue$(s)$ reaches
line~\ref{li:ifallred}, $s$ must still be cyan, and its colour is
changed by of the statements in lines \ref{li:makered}, \ref{li:makered3},
or \ref{li:makeblue} to either red or blue.

\paragraph{Meaning of colours}
From the above, we can deduce the following:
\begin{itemize}
\item A state is white if and only if it has never been touched by \dfsblue.
\item A state is cyan if and only if its invocation of \dfsblue{} is still
 running, (i.e., it is on the ``search stack'' of \dfsblue), and every cyan
 state can reach $s$, if \dfsblue$(s)$ is the currently active instance
 of \dfsblue.
\item A state is blue if and only if it is non-accepting and its invocation
 of \dfsblue{} has terminated.
\item If a state is red, its invocation of \dfsblue{} has terminated, and it
 is not part of any counterexample.
\end{itemize}
The last part of this statement is proved in the next paragraph.

\paragraph{Red states}
We prove that red states are never part of any counterexample. More
precisely, whenever an invocation of \dfsblue{} terminates,
all states that have been coloured red by that time are not part of any
counterexample. We proceed by induction on the states in
the post-order implied by \dfsblue, or, put differently, we show that
this property is an invariant of the program.

Obviously, the statement holds initially because there are no red states.
Now, suppose that some state~$s$ is made red by line~\ref{li:makered}.
Then, all its successor states are red, so by induction hypothesis none
of them are part of any counterexample. Since any counterexample including~$s$
also has to include one of its successors, $s$ cannot be part of a
counterexample.

It remains to show that lines \ref{li:redcall} and~\ref{li:makered3} preserve
the invariant. Assume therefore that the call to \dfsred{} in
line~\ref{li:redcall} terminates. We now show that in this case, no
state~$s'$ visited by \dfsred{} is part of any counterexample. Assume
by contradiction that $s'$ is part of a counterexample. Then there
must be some accepting state~$t$ reachable from $s'$ (and therefore from~$s$),
and there must be a path from $s$ via $s'$ to $t$ in which all states were
non-red before line~\ref{li:redcall} was reached (by induction hypothesis,
because these states are part of a counterexample). However, such a state~$t$
cannot exist:
\begin{itemize}
\item $t$ cannot be white because it is reachable from $s$, and therefore it
 must have been visited by \dfsblue{} before \dfsblue$(s)$ could have reached
 line~\ref{li:ifallred}.
\item $t$ cannot be cyan because it is reachable from $s$ by non-red states,
 and therefore \dfsred{} would terminate when reaching $t$.
\item $t$ cannot be blue because it is accepting.
\item $t$ cannot be red because this means that its invocation of \dfsblue{}
 has already finished, in which case, by induction hypothesis, it is not
 part of any counterexample.
\end{itemize}

\paragraph{Correctness, part 1}
We now show that whenever the algorithm reports a cycle, a counterexample
indeed exists. Cycles are reported in lines \ref{li:bluecycle}
and~\ref{li:redcycle}. 

In line~\ref{li:bluecycle}, there is a transition from $s$ to~$t$. Since $t$
is cyan, there is also a path from $t$ to $s$, and either $s$ or $t$ are
accepting. Therefore, a counterexample exists.

In line~\ref{li:redcycle}, there is a transition from $s$ to~$t$.
Assume that $s'$ is the ``seed'' of the current red DFS, i.e.\ $s'$ was
the state that most recently reached line~\ref{li:redcall}. Then, $s'$
is accepting and can reach~$s$. Moreover, since $t$ is cyan, it can
reach $s'$, completing the counterexample.

\paragraph{Correctness, part 2}
We now show that whenever a counterexample exists, the algorithm reports one.
Let $s$ be an accepting state within the loop of such a counterexample. Then,
either the algorithm reaches line~\ref{li:redcall} in the \dfsblue{} invocation
on~$s$, or it will terminate even earlier with a counterexample. We show that
in the first case the red DFS on $s$ will still find a counterexample.

Consider the states forming the loop of the counterexample at the time
when \dfsred$(s)$ is called. None of them can be red, and none of them
can be white because they are all reachable from~$s$ and therefore have
been considered by \dfsblue{} earlier. This, all of them are either
blue or cyan. In particular, at least one state in the loop, i.e., $s$ itself,
is still cyan. Therefore, the red search is guaranteed
to find a cyan state and report a counterexample.

\section{SCC-based algorithms}
\label{sec:scc}

An efficient algorithm for determining SCCs that works on-the-fly was
first proposed by Tarjan~\cite{Tar72}. However, for model-checking purposes
Tarjan's algorithm was deemed unsuitable because it used more memory than
nested DFS while offering no advantages. More recent innovations by
Geldenhuys/Valmari~\cite{GV04} and Couvreur~\cite{Cou99} change the
picture, however: their modifications allow SCC-based analysis to report
a counterexample as soon as all its states and transitions were discovered,
no matter in which order. In other words, if the order in which successors
are explored by the DFS is fixed, both can find a counterexample in optimal
time (w.r.t.\ to the exploration order).

Space constraints prevent us from presenting the algorithms in detail.
However, we mention a few salient points. Tarjan places all
newly discovered states onto a stack (henceforth called \emph{Tarjan stack})
and numbers them in pre-order. Certain properties of the DFS ensure
that at any time during the algorithm, states belonging to the same SCC are
stored consecutively on the stack and therefore also numbered consecutively.
The \emph{root} of an SCC is the state explored first during DFS, having the
lowest number and being deepest on the Tarjan stack.
For each state~$s$, Tarjan computes a so-called ``lowlink'' number, which
is identical to the number of $s$ iff $s$ is a root, and less
than that otherwise. An SCC is completely explored when backtracking
from its root, and at that point it can be identified as a complete SCC
and removed from the Tarjan stack.

Geldenhuys/Valmari (GV) exploit properties of lowlinks; they remember the
number of the deepest accepting state on the current search path, say $k$,
and when a state with lowlink $\le\;k$ is found, a counterexample is
reported. They also propose some memory savings that
are of minor importance in our context.

Couvreur (C99) omits both Tarjan stack and lowlinks but
introduces a \emph{roots stack} that stores the roots of
all partially explored SCCs on the current search path. When one finds
a transition to a state with number $k$, properties of the numbering imply that
no state with number larger than $k$ can be a root,
prompting their removal from the roots stack. This 
effectively merges some SCCs, and one checks whether the
merger creates an SCC with the conditions from Fact~(2).

Both algorithms report a counterexample after seeing the same states and
transitions, provided they work with the same exploration order. However,
it turns out that the removal of the Tarjan stack in C99,
while more memory efficient, was a crucial oversight:
when
backtracking from a root, another DFS is necessary to mark these states
as ``removed''. These extra $\post$ computations severely impede its
performance. This makes GV superior to C99 in practice.

We propose to amend C99 by re-inserting the Tarjan
stack.\footnote{The problem with C99 was first hinted at in~\cite{SE05}.
After creating this improvement independently,
we learned that similar changes were already proposed in \cite{CDP05}
and \cite{GV05}.}
This amendment makes it competitive with GV while using
slightly less memory; more crucially, C99 can deal directly with GBAs,
which GV cannot. Since GBAs tend to be smaller than BAs for the same
LTL formula, the amended algorithm can hope to explore fewer states
and be faster.


\begin{figure}[t]
\begin{center}
\begin{minipage}[t]{5.5cm}
{\algo
[procedure] {\it couv} ()
>${\it count} := 0$;
>${\it Roots} := \emptyset$; ${\it Active} := \emptyset$;
>[call] {\it couv\_dfs}($s_I$)}

\medskip

{\algcontd
[procedure] {\it couv\_dfs}($s$):
>${\it count} := {\it count} + 1$;		\label{li:startini}
>$s.{\it dfsnum} := count$;
>$s.{\it current} := [true]$;
>[push]$({\it Roots},(s,A(s)))$;
>[push]({\it Active},s);			\label{li:endini}
>\for{t\in\post(s)              		\label{li:forloop}}
>>\ifthen{t.{\it dfsnum}=0}
>>>[call] {\it couv\_dfs}($t$)       		\label{li:fresh}}
\end{minipage}
\begin{minipage}[t]{6.5cm}
{\algcontd
>>[else] \ifthen{t.{\it current}}		\label{li:removed}
>>>$B:=\emptyset$;
>>>[repeat]                   		 	\label{li:collapse}
>>>>$(u,C) := [pop]({\it Roots})$;
>>>>$B:=B\cup C$;				\label{li:union}
>>>>\ifthen{B=K} [report cycle]      		\label{li:cycle}
>>>[until] $u.{\it dfsnum}\le t.{\it dfsnum}$;
>>>[push](${\it Roots},(u,B)$);             	\label{li:pushz}
>\ifthen{[top]({\it Roots})=(s,?)}    		\label{li:top}
>>[pop]({\it Roots});
>>[repeat]
>>>u:=[pop]({\it Active});
>>>u.{\it current} := [false]
>>[until] $u=s$}
\end{minipage}
\end{center}
\caption{Amendment of Couvreur's algorithm.}
\label{fig:newscc}
\end{figure}

The amended algorithm, working with GBAs, is shown in Figure~\ref{fig:newscc},
and a proof of correctness is given below.
Note that in C99 accceptance conditions are annotated on the transitions,
whereas here we place them on the states, which is only a minor
difference. Figure~\ref{fig:newscc} assumes $k$ acceptance sets,
denoting $\A(s):=\{\,j\mid s\in A_j\,\}$ and $K:=\{1,\ldots,k\}$.
Note that if $k$ is ``small'', the union operation in line~\ref{li:union}
can be implemented with bit parallelism.

\subsection{Proof of the new algorithm}

We provide a detailed proof of correctness. The proof works from scratch
and assumes only very basic knowledge of graph theory plus the concept
of SCCs.

\def\num{\mathit{num}}
\def\cdfs{{\it couv\_dfs}}
\def\A{\mathcal{A}}
\def\E{\mathcal{E}}

\paragraph{Basic Definitions}
A \emph{DFS numbering} of $\B$ is a pre-order numbering starting at
the initial state~$s_I$. In general, depending on the order in which
successors are explored, an automaton has many possible DFS numberings;
here we assume one externally fixed order and therefore one fixed
DFS numbering. The number assigned to state~$s$ is denoted $\num(s)$.
Note that states are added to the Tarjan stack (called $\mathit{Active}$
in Figure~\ref{fig:newscc}) in the order of their numbering.

The \emph{root} of an SCC within $\B$ is the state visited first
by \cdfs{} during the algorithm. (Precisely which state within an SCC
is a root may also depend on the exploration order.)

At any time during the algorithm, we mean by \emph{search path} the
sequence of currently unfinished calls to \cdfs.

\paragraph{Subgraphs of~$\B$}
A state~$s$ is called \emph{explored} when \cdfs$(s)$ has been
called. A transition from $s$ to $t$ is called \emph{explored}
when $t$ appears in the for-loop during execution of \cdfs$(s)$.
At any time during the algorithm, we mean by \emph{explored graph}
the subgraph~$\E$ consisting of all explored states and transitions.

We call an SCC of~$\E$ \emph{active} if the search
path contains at least one of its states. Note that
the SCCs of $\E$ may be different from those of~$\B$!
In particular, due to unexplored transitions, two SCCs
of~$\E$ may be part of the same SCC of~$\B$.

A state is called \emph{active} if it is part of an active SCC.
The state itself need not be on the search path.

At any time during the algorithm, we mean by \emph{active graph}
the subgraph~$\A$ induced by the active states.

\paragraph{Facts}
\begin{enumerate}
\item \label{fact:searchpath}
  Let $s_0\cdots s_n$ be the search path at any time.
  Then $\num(s_i)\le\num(s_j)$ iff $i\le j$.
  Moreover, $s_i\reach s_j$ if $i\le j$.
  \\
  Proof: immediate from the logic of the program.

\item \label{fact:rootnum}
  A root has the least number and lies lowest on $\mathit{Active}$
  within its SCC.
  \\
  Proof: obvious

\item  \label{fact:rootlast}
  Within each SCC, the root is the last state from which \cdfs{}
  backtracks, and at that point, the SCC has been completely explored
  (i.e., all states and edges have been considered).
  \\
  Proof: Suppose \cdfs{} reaches root $r$ of SCC~$C$.
  At that point, no other state of~$C$ has been visited so far,
  and all are reachable from~$r$. Therefore, the DFS will visit
  all those states (and possibly others) and backtrack from them
  before it can backtrack from~$r$.

\item \label{fact:backtrack}
  An SCC becomes inactive when we backtrack from its root.
  \\
  Proof: follows from Fact~\ref{fact:rootlast}.

\item \label{fact:inactive}
  An inactive SCC of~$\E$ is also an SCC of~$\B$.
  \\
  Proof: follows from Facts \ref{fact:rootlast} and~\ref{fact:backtrack}.

\item \label{fact:subseq}
  The roots of~$\A$ are a subsequence of the search path.
  \\
  Proof: follows from Fact~\ref{fact:backtrack} because the root of
  an active SCC must be on the search path.

\item \label{fact:uniqueroot}
  Let $s$ be an active state and $t$ the root of its SCC in~$\A$.
  Then $\num(t)\le\num(s)$ and there is no active root $u$
  with $\num(t)<\num(u)<\num(s)$.
  \\
  Proof: The first part is just a consequence from Fact~\ref{fact:rootnum}.
  For the second part, assume that such an active root~$u$ exists.
  Since $u$ is active, it is on the search stack, just like~$t$,
  which follows from Fact~\ref{fact:subseq}. From Fact~\ref{fact:searchpath},
  we have $t\reach u$. As \cdfs$(u)$ has not yet terminated and
  $\num(u)<\num(s)$, $s$ must have been reached from~$u$,
  i.e.\ $u\reach s$. Because $s$ and $t$ are in the same SCC,
  $s\reach t$. But then, $t$ and $u$ are in the same SCC
  and cannot both be its root.

\item \label{fact:numreach}
  Let $s$ and $t$ be two active states with $\num(s)\le\num(t)$.
  Then $s\reach t$.
  \\
  Proof: Let $s',t'$ be the (active) roots for $s$ and $t$, resp.
  From Fact~\ref{fact:uniqueroot} we have $\num(s')\le\num(t')$,
  thus from Fact~\ref{fact:searchpath} we have
  $s'\reach t'$, and therefore $s\reach t$.
\end{enumerate}

\paragraph{Conclusions}

From the facts that we have just shown, we can conclude that the active
graph~$\A$ always has the kind of shape visualized in
Figure~\ref{fig:invariant}, with the following properties:

\begin{figure}
\centerline{\includegraphics[width=12cm]{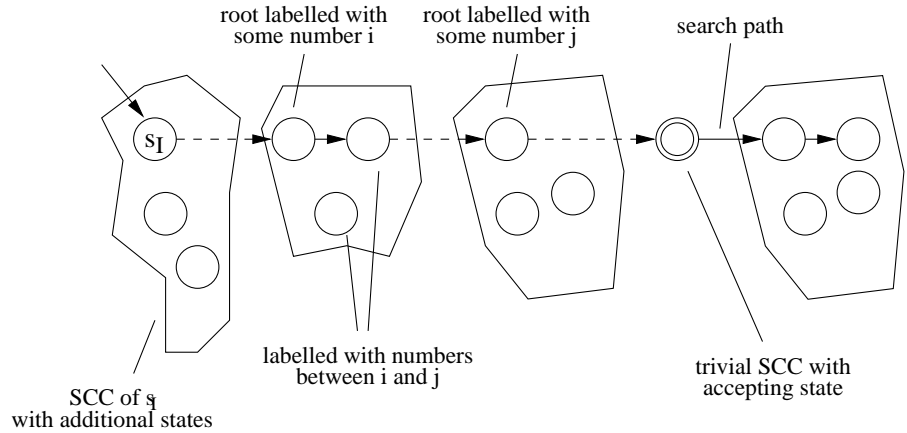}}
\caption{Shape of the active graph}
\label{fig:invariant}
\end{figure}

\begin{itemize}
\item The search path (indicated by the connected line of states at the top
  of the figure) is completely contained in the active graph, and its roots
  from a subsequence of the search path.
\item The SCCs are ``linearly ordered'', i.e. if one defines $C_1<C_2$
  iff $C_2$ can be reached from $C_1$, then $<$ is a total order.
\item The DFS numbering is consecutive in the sense that if $i$ and $j$
  are the numbers of two subsequent roots on the search path, then the
  active states with numbers $n$ such that $i\le n<j$ form an SCC. From
  this it follows that these states are also consecutive on the Tarjan stack.
\end{itemize}

\paragraph{Correctness of the algorithm}
The correctness of the algorithm is now easy to show. We assume that all
newly discovered states are initialized with a number~$0$ and a false
$\mathit{current}$ bit. It then suffices to prove
that the algorithm maintains the following invariants after each exploration
of a state or transition:
\begin{itemize}
\item The $\mathit{Roots}$ stack contains the roots of the active graph
   (in the order implied by the search path) together with the union
   of all acceptance indices occurring within the corresponding SCC of~$\A$.
\item The $\mathit{Active}$ stack contains exactly the active states, and
   exactly the active states have the $\mathit{current}$ bit set to true.
\end{itemize}

In the beginning of the algorithm, this invariant holds because the active
graph contains just $s_I$ and no transitions. Thus, the single
element of $\mathit{Roots}$ is $(s_I,A(s_I))$, and $s_I$ is active.
This is ensured by the first part up to line~\ref{li:endini}.

The invariant is then upheld whenever a transition from some $s$ to some $t$
is discovered.  There are five cases:
\begin{itemize}
\item $t$ is a newly discovered state. In this case, $\A$ is extended
  by $t$ and the transition $s\to t$, and $t$
  forms a new trivial SCC within~$\A$. No counterexample is generated
  in this way. The recursive call in line~\ref{li:fresh}
  and lines \ref{li:startini} through \ref{li:endini} perform the
  necessary actions.
\item $t$ has been visited before and is inactive. Then, its SCC has
  been completely explored, so $s$ and $t$ belong to different SCCs,
  so $t\not\reach s$. The edge $s\to t$ cannot be part of a loop,
  the active graph does not change, so no action is necessary.
\item $t$ is active and $\num(t)>\num(s)$.
  From Fact~\ref{fact:numreach} we already know that $s\reach t$ holds,
  therefore
  this discovery does not change the SCCs and no new counterexample can be
  generated in this way. Thus, no action is necessary.
\item $t$ is active and $\num(t)=\num(s)$.
  Then $s=t$, and a counterexample has been discovered iff $s$ contains
  all acceptance conditions. Otherwise, the SCCs of the active graph
  remain unchanged.
\item $t$ is active and $\num(t)<\num(s)$.
  Then from Fact~\ref{fact:numreach} we know $t\reach s$, so $s$ and $t$
  belong to the same SCC. Let $u$, with $\num(u)\le\num(t)$ be
  the root of the SCC to which $t$ belongs. Since $s$ is the last
  element on the search path, it follows from Fact~\ref{fact:searchpath}
  that all SCCs on the $\mathit{Roots}$ stack from $u$ downwards
  must be merged into one SCC. Moreover, $u$ is the unique topmost
  root on $\mathit{Roots}$ whose number is no larer than
  $\num(t)$ according to Fact~\ref{fact:uniqueroot}. Finally,
  the merger yields a non-trivial SCC, and
  a new counterexample is generated iff the SCC
  contains all acceptance conditions.
\end{itemize}
The last three cases are dealt with uniformly in lines
\ref{li:removed} through \ref{li:pushz} of Figure~\ref{fig:newscc}.

Finally, when backtracking from a state $s$, two cases can happen:
\begin{itemize}
\item $s$ is a root. Then necessarily the $\mathit{Roots}$ stack
  has a topmost entry with $s$ because $s$ is currently last
  on the search path, and said entry must be removed.
  Moreover, the entire SCC becomes inactive according
  to Fact~\ref{fact:backtrack}. This is dealt with from
  line~\ref{li:top} downwards.
\item $s$ is not a root. Then the topmost $\mathit{Roots}$
  entry does not show~$s$, no node becomes inactive, and
  no further action is necessary.
\end{itemize}
Thus, the invariant is upheld. A counterexample is reported as
soon it is contained in the explored graph~$\E$. As a consequence,
if the algorithm terminates normally, no counterexample exists.

\section{Experiments}
\label{sec:experiments}

We implemented a framework for testing and comparing the actual
performance of all the known B\"uchi emptiness algorithms.
For practical relevance, the best
framework for such an implementation would have been Spin. However, 
Spin turned out too difficult to modify for this purpose. Instead, we
based our testbed on NIPS~\cite{Web07}, a reverse-engineered Promela engine.
Essentially, NIPS allows to process a Promela model, provides the initial
state descriptor and a function for computing its successors. It is thus
ideally suited for testing on-the-fly algorithms, and we believe that the
conditions are as close to Spin as possible.

We used 266 test cases from the BEEM database~\cite{Pel07}, including
many different algorithms, e.g., the Sliding Window protocol, Lamport's
Bakery algorithm, Leader Election, and many others, together with various
LTL properties.

Among the algorithms tested and implemented were HPY~\cite{HPY96},
GV~\cite{GV04}, C99~\cite{Cou99}, SE~\cite{SE05},
and the amended algorithms presented
in Sections~\ref{sec:nested} and~\ref{sec:scc}, henceforth called AND
and ASCC. For weak automata, we report on simple DFS (SD,
see Section~\ref{sec:nested}).
We also implemented and tested other algorithms, notably those
from \cite{CVWY92} and \cite{GMZ04}.
However, these were always
dominated by others, and we omit them in the following.
Naturally, our concrete running times and memory consumptions are subject
to certain implementation-specific issues. Nonetheless, we believe that the
tendencies exhibited by our experiments are transferrable.

In the following, we give a summary of our results. A more detailed description
of our framework, the benchmarks, and the experimental results is given
in~\cite{Gai07}; here, we just summarize the most important findings.

We first found that, in the context of exact model checking, the
differences in auxiliary memory usage was basically irrelevant.
Certainly, the auxiliary memory used by the various algorithms ranged
from 2 bits to 12 bytes, a comparatively large difference. However, this
was dwarved by the memory consumption of state descriptors, which ranged
from 20 to 380 bytes, averaging at 130.

The only practical difference therefore was in the running time. Here, we
found that, no matter which auxiliary data structures were employed, the
running time was practically proportional to the number of $\post$ invocations
(more precisely: the number of individual successor states generated by
$\post$),
by far the most costly operation. In retrospect, these two
observations may seem obvious; however, we find that they were consistently
under-represented in previous papers, therefore it is worth re-emphasizing
their relevance.
The two main factors contributing to the running time were fast counterexample
detection and whether an algorithm had to compute each transition at most once
or twice.

Discussing individual test cases would not be very meaningful: for instance,
the early-detection properties of some algorithms can cause arbitrarily
large differences. Instead, we exhibit certain structural properties
that occurred in many test cases and caused those differences. We first discuss
algorithms working on ``normal'' B\"uchi automata, followed by a discussion
of ASCC with GBAs.

First, we observe that most test cases constitute weak B\"uchi automata.
Note that the intersection BA is weak if the BA arising from the formula is
weak. \v{C}ern\'a and Pel\'anek~\cite{CP03} estimate the proportion of
weak formulae in practice to 90--95~\%; indeed, we found that only 8~\% of
our test cases were non-weak.
For weak test cases, five out of six tested
algorithms (GV, C99, SE, AND, SD) detect counterexamples
with minimal exploration. The three main structural effects causing performance
differences (which may overlap) were as follows:
\begin{itemize}
\item In 86 test cases, we observed many trivial SCCs consisting of one
 accepting state. A typical example is the LTL property $GF p$, which
 (when negated) yields a weak automaton with a looping accepting state.
 Then, any non-looping part of the system necessarily yields such trivial SCCs.
 In these cases, GV and SD dominate, sometimes with a factor of two, whereas
 C99, SE, and HPY fall behind because
 they explore the accepting trivial SCCs twice. In our test cases, the AND
 algorithm had the same performance as GV and SD, although this is not
 guaranteed in general.
\item In 98 cases, we observed non-accepting SCCs not preceded
 by accepting SCCs. In this case, C99 falls behind all the others.
\item HPY reports counterexamples only during the red DFS, whereas SE and AND
 discovers some during the blue DFS. This accounts for 101 test cases in which
 HPY fared worst, whereas all others showed the same performance.
\end{itemize}

\begin{floatingfigure}[r]{4cm}
\begin{center}
\begin{tabular}{l@{\ \ }|@{\ \ }r}
algorithm & run-time \\
\hline
ASCC & 67.0~\% \\
GV   & 69.2~\% \\
AND  & 69.7~\% \\
SE   & 96.3~\% \\
HPY  & 100.0~\% \\
C99  & 128.3~\%\\
\end{tabular}
\end{center}

\caption{Performances}
\label{fig:results}
\end{floatingfigure}

Non-weak automata also had these effects, affecting
18, 17, and 7 out of 21 test cases. In 7 cases,
GV and C99 found counterexamples more quickly than the others,
being faster by
a factor of up to six. Since we used the same exploration order in
all algorithms, these results are directly comparable.

We then tested the ASCC algorithm with GBA, generated
by the LTL2BA tool~\cite{GO01}. Most formulae yielded GBA with only one
acceptance condition, meaning that the GBA had the same size as the
corresponding BA. 
Notice that the running times of GBA with multiple conditions are not directly
comparable with those of the corresponding BA. This is because using a different
automaton changes the order of exploration, therefore in some ``lucky'' cases
the BA-based algorithms may still find a counterexample more quickly.

The running times summed up over all 266 test cases are given in
Figure~\ref{fig:results}, expressed as percentages of each other.
Additionally, SD had the same performance
as GV for the weak cases.
Note that every set of benchmarks would lead to the same order
among the algorithms because it reflects their different qualitative
properties
(e.g., quick counterexample detection or number of $\post$ calls).
The concrete numbers in Figure~\ref{fig:results} tell their quantitative
effect in what we believe to be a representative
set of benchmarks.
We draw the following conclusions:
\begin{itemize}
\item Because of the dominance of weak test cases and GBAs with only one
  acceptance condition, the sum of running times yields small differences;
  only SE, HPY, and C99 clearly fall behind. The performance differences in the
  comparatively few other cases is very pronounced however.
\item Overall, ASCC is the best algorithm if GBAs can be used. Due to the
  technical reasons explained above, it did not perform best in all examples.
\item Among the BA-based algorithms, GV is the best for general formulae;
  it is never outperformed on any test case by any other BA-based algorithm.
  ASCC performs equally well when used with simple BAs.
\item For weak formulae, SD is the best algorithm for
  bitstate hashing.
\item For general formulae, AND is the best algorithm for
  bitstate hashing, improving the previous best algorithm
  for this setting (SE) by 28~\%.
\item There remains no reason to use either SE, HPY, or C99.
\end{itemize}

\subsubsection{Acknowledgements:}
The authors would like to thank Michael Weber
for creating and helping us use the NIPS framework.

\bibliographystyle{plain}
\bibliography{db}

\begin{thebibliography}{10}

\bibitem{BBR08}
Ji\v{r}\'{\i} Barnat, Lubo\v{s} Brim, and Petr Ro\v{c}kai.
\newblock {DiVinE} multi-core - a parallel {LTL} model-checker.
\newblock In {\em Proc.\ ATVA}, LNCS 5311, pages 234--239, 2008.

\bibitem{CVWY92}
Costas Courcoubetis, Moshe~Y. Vardi, Pierre Wolper, and Mihalis Yannakakis.
\newblock Memory-efficient algorithms for the verification of temporal
  properties.
\newblock {\em Formal Methods in System Design}, 1(2/3):275--288, 1992.

\bibitem{Cou99}
Jean-Michel Couvreur.
\newblock On-the-fly verification of linear temporal logic.
\newblock In {\em Proc.\ Formal Methods}, LNCS 1708, pages 253--271, 1999.

\bibitem{CDP05}
Jean-Michel Couvreur, Alexandre Duret-Lutz, and Denis Poitrenaud.
\newblock On-the-fly emptiness checks for generalized {B\"uchi} automata.
\newblock In {\em Proc.\ SPIN}, LNCS 3639, pages 169--184, 2005.

\bibitem{FFK+01}
Kathi Fisler, Ranan Fraer, Gila Kamhi, Moshe~Y. Vardi, and Zijiang Yang.
\newblock Is there a best symbolic cycle-detection algorithm?
\newblock In {\em Proc.\ TACAS}, LNCS 2031, pages 420--434, 2001.

\bibitem{Gai07}
Andreas Gaiser.
\newblock {Vergleich von Algorithmen f\"ur den Leerheitstest von
  B\"uchiautomaten}.
\newblock Studienarbeit, Universit\"at Stuttgart, 2007.
\newblock In German.

\bibitem{GS09}
Andreas Gaiser and Stefan Schwoon.
\newblock Comparison of algorithms for checking emptiness on {B\"uchi}
  automata.
\newblock In {\em Proc.\ MEMICS}, 2009.

\bibitem{GMZ04}
Paul Gastin, Pierre Moro, and Marc Zeitoun.
\newblock Minimization of counterexamples in {SPIN}.
\newblock In {\em Proc.\ 11th SPIN Workshop}, LNCS 2989, pages 92--108, 2004.

\bibitem{GO01}
Paul Gastin and Denis Oddoux.
\newblock Fast {LTL} to {B}{\"u}chi automata translation.
\newblock In {\em Proc.\ CAV}, LNCS 2102, pages 53--65, 2001.

\bibitem{GV04}
Jaco Geldenhuys and Antti Valmari.
\newblock {Tarjan's} algorithm makes on-the-fly {LTL} verification more
  efficient.
\newblock In {\em Proc.\ TACAS}, LNCS 2988, pages 205--219, 2004.

\bibitem{GV05}
Jaco Geldenhuys and Antti Valmari.
\newblock More efficient on-the-fly {LTL} verification with {Tarjan's}
  algorithm.
\newblock {\em Theoretical Computer Science}, 345(1):60--82, 2005.

\bibitem{Hol03}
Gerard~J. Holzmann.
\newblock {\em The {Spin} Model Checker: Primer and Reference Manual}.
\newblock Addison-Wesley, 2003.

\bibitem{HPY96}
Gerard~J. Holzmann, Doron~A. Peled, and Mihalis Yannakakis.
\newblock On nested depth first search.
\newblock In {\em Proc.\ 2nd SPIN Workshop}, pages 23--32, 1996.

\bibitem{Pel07}
Radek Pel{\'a}nek.
\newblock Beem: Benchmarks for explicit model checkers.
\newblock In {\em Proc.\ SPIN}, LNCS 4595, pages 263--267, 2007.

\bibitem{SE05}
Stefan Schwoon and Javier Esparza.
\newblock A note on on-the-fly verification algorithms.
\newblock In {\em Proc.\ TACAS}, LNCS 3440, pages 174--190, 2005.

\bibitem{Tar72}
Robert Tarjan.
\newblock Depth-first search and linear graph algorithms.
\newblock {\em SIAM Journal on Computing}, 1(2):146--160, 1972.

\bibitem{Tau06}
Heikki Tauriainen.
\newblock Nested emptiness search for generalized {B\"uchi} automata.
\newblock {\em Fundamenta Informaticae}, 70(1--2):127--154, 2006.

\bibitem{CP03}
Ivana \v{C}ern\'a and Radek Pel\'anek.
\newblock Relating hierarchy of linear temporal properties to model checking.
\newblock In {\em Proc.\ MFCS}, LNCS 2747, pages 318--327, 2003.

\bibitem{Web07}
Michael Weber.
\newblock An embeddable virtual machine for state space generation.
\newblock In {\em Proc.\ SPIN}, LNCS 4595, pages 168--186, 2007.

\end{thebibliography}

\end{document}